\documentclass[]{spie}  
\usepackage[]{graphicx}
\usepackage[]{epstopdf}
\title{Experimental parametric study of the Self-Coherent Camera} 

\author{Johan Mazoyer\supit{a}, Pierre Baudoz\supit{a}, Marion Mas\supit{a}\supit{b}, Gerard Rousset\supit{a},Rapha\"el Galicher\supit{c}\supit{d}
\skiplinehalf
\supit{a}LESIA, Observatoire de Paris, CNRS, Universit\'e Pierre et Marie Curie Paris 6 and Universit\'e Denis
Diderot Paris 7,  5 place Jules Janssen, 92195 Meudon, France; \\
\supit{b}Laboratoire d'Astrophysique de Marseille, CNRS-INSU, Universit\'e d’Aix-Marseille, 38 rue Fr\'ed\'eric Joliot-Curie, 13388 Marseille cedex 13 France; \\
\supit{c}National Research Council Canada, Herzberg Institute of Astrophysics, 5071 West Saanich Road,
Victoria, BC, V9E 2E7, Canada; \\
\supit{d}Dept. de Physique, Universit\'e de Montr\'eal, C.P. 6128 Succ. Centre-ville, Montr\'eal, Qc, H3C 3J7, Canada; \\
}

\authorinfo{Further author information: (Send correspondence to Johan Mazoyer)\\Johan Mazoyer E-mail: johan.mazoyer@obspm.fr, Telephone: +33 (0)1 98 76 54 32}

\begin{document} 
  \maketitle 

\begin{abstract}
Direct imaging of exoplanets requires the detection of very faint objects orbiting close to very bright stars. In this context, the SPICES mission was proposed to the European Space Agency for planet characterization at visible wavelength. SPICES is a 1.5m space telescope which uses a coronagraph to strongly attenuate the central source. However, small optical aberrations, which appear even in space telescopes, dramatically decrease coronagraph performance. To reduce these aberrations, we want to estimate, directly on the coronagraphic image, the electric field, and, with the help of a deformable mirror, correct the wavefront upstream of the coronagraph. We propose an instrument, the Self-Coherent Camera (SCC) for this purpose. By adding a small ``reference hole" into the Lyot stop, located after the coronagraph, we can produce interferences in the focal plane, using the coherence of the stellar light. We developed algorithms to decode the information contained in these Fizeau fringes and retrieve an estimation of the field in the focal plane. After briefly recalling the SCC principle, we will present the results of a study, based on both experiment and numerical simulation, analyzing the impact of the size of the reference hole.
\end{abstract}



\section{INTRODUCTION}
\label{sec:intro}  

Since the first detection of an exoplanet around a solar type star in 1995, the search and analysis of such planets have been very active fields of research, using both theoretical models and space and ground observations. Most of these planetary systems have been discovered by indirect methods such as transits and radial velocities, analyzing only the light from the star. These methods are well complemented by direct imaging. Not only because it gives us a chance to reach different types of planets (further from their star), but also because it allows a spectral study of the planet light. Spectroscopy is essential for the complete characterization of the planet, as we gain access to the chemistry of its atmosphere and to the surface pressure and temperature. In order to achieve direct imaging of faints objects orbiting close to very bright stars, it is necessary to reduce the light coming from the star by many orders of magnitude.

Different ground-based instruments are now being developed to reach the high contrasts of $10^{-6}$ or $10^{-7}$ which are necessary to analyze hot Jupiter-size planets. A lot of these direct imaging methods use coronagraphs. But due to the initial aberrations in the wave front, stellar leakage speckles remain in the focal plane even after the coronagraph. In order to study smaller and colder objects such as Neptune-like or even rocky planets, an important improvement of these contrasts is necessary. The use of larger mirrors, as in the next generation of Extremely Large Telescopes (ELT) or of dedicated embedded instruments, will not achieve these contrasts without a better understanding of the fundamental limitations inherent to this technique. In this context, the Self-Coherent Camera\cite{Baudoz06}, a solution to reduce the intensity of these speckles has been developed at the Laboratoire d'Etude Spatiales et d'Instrumentation en Astrophysique (LESIA), at Paris Observatory. This instrument uses the properties of coherence between the speckles, leaks of the star light in the focal plane, and the light rejected by the coronagraphic mask. It can be used either to estimate the phase aberrations\cite{Galicher10} or to detect planets among speckles\cite{BaudozSPIE12}. The SCC can be associated with most coronagraphs and this technique is thus a good candidate for new generation ground telescopes as well as for future space missions.\cite{Galicher10,Maire12}. 

After recalling the principle of the Self-Coherent Camera in Section \ref{II}, we will analyze the influence of the size of the reference pupil on the performances of this instrument, using numerical simulations as well as results obtained on the optical test bench described in Section \ref{II4}.

\section{SCC description and role of the reference pupil} 
\label{II}

\subsection{Principle of the Self-Coherent Camera} 
\label{II1}

\begin{figure}
   \begin{center}
   \begin{tabular}{c}
    \includegraphics[width=10.5cm]{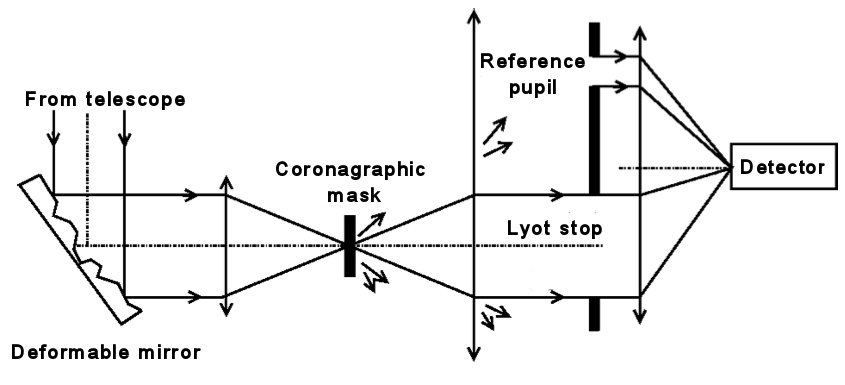}
   \includegraphics[width=5.5cm]{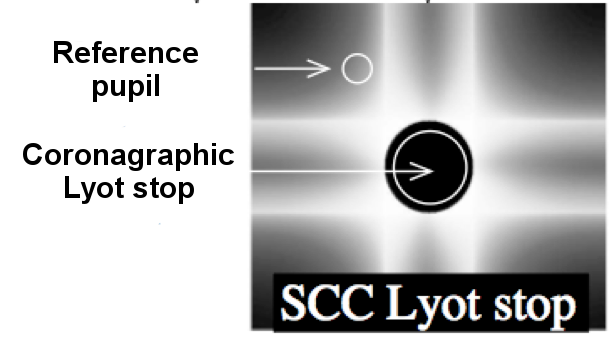}
 
   \end{tabular}
   \end{center}
   \caption[Principle of the SCC] 
   { \label{fig:SCC_princ} 
Principle of the SCC in association with a coronagraph and a deformable mirror (left) and light distribution in the pupil plane downstream of a Four Quadrant Phase Mask (FQPM) coronagraph\cite{Rouan00} }
   \end{figure}

In a standard coronagraph, the light of an on-axis source, after the entrance pupil, is focalized on a focal plane mask, which diffracts it outside of the pupil geometry. A Lyot stop, introduced in the downstream pupil plane, removes the diffracted light. When imaging a planetary system, we can theoretically completely suppress the light of the star and disclose the companion in the detector plane. But small aberrations, inherent to the instrument optics, produce leaks of the stellar light in the detector plane, also called speckles, which are the same size and can be brighter than planets images. The goal of the SCC is to discriminate the residuals of the stellar light from the light of a companion. The principle of the SCC is described in Figure \ref{fig:SCC_princ} (left). Only a simple modification (Figure \ref{fig:SCC_princ}, right) of the Lyot stop is necessary to implement the SCC in a coronagraph. By adding an off-axis small pupil in the lyot stop plane which we call the ``reference pupil", we are able to select only stellar light diffracted by the coronagraph. The speckles from the stellar leakage will interfere with the light coming from the reference pupil and will be fringed. This encoding will allow us to estimate the upstream phase and amplitude aberrations, as shown in Section \ref{II2}. From the measurement of these aberrations, we are able to correct them using a deformable mirror, placed upstream of the coronagraph. Practically, this correction will imply the building of an interaction matrix, by applying a basis of known aberrations on the deformable mirror and observing its effect on our phase estimator. To mitigate noise effects, we will then construct a synthetic matrix. Inverting this matrix will allow us to correct unknown aberrations. For more information about this process, see Mas et al. 2012\cite{MasSPIE12}.

\subsection{Principle of phase estimation} 
\label{II2}

As already described in a previous work (Galicher et al. 2010\cite{Galicher10}), the image on the detector plane (after the coronagraph and SCC) can be written at wavelength $\lambda_{0}$ as:

\begin{equation}
	\label{eq:I_in_f_plan}
\ I(\alpha) = |A_{S}(\alpha)|^2 + | A_{R}(\alpha)|^2 + A_{S}(\alpha)A_{R}^{*}(\alpha)\exp\left( \frac{2i\pi\alpha\xi_{0}}{\lambda_{0}} \right) + A_{S}^{*}(\alpha)A_{R}(\alpha) \exp\left( -\frac{2i\pi\alpha\xi_{0}}{\lambda_{0}} \right),
	\end{equation}
where $A_{S}(\alpha)$ (respectively $A_{R}(\alpha)$) is the complex amplitude in the focal plane after propagation through Lyot pupil (respectively reference pupil). This function is the Fourier transform of $\psi_{S}$ (respectively $\psi_{R}$), the complex amplitude in the pupil plane downstream of the Lyot stop (respectively, downstream of the reference pupil). $\alpha$ is the focal plane angular coordinate, and $\xi_{0}$ is the separation between the Lyot stop and the reference pupil.

\begin{figure}
   \begin{center}
   \begin{tabular}{c}
   \includegraphics[width=8cm]{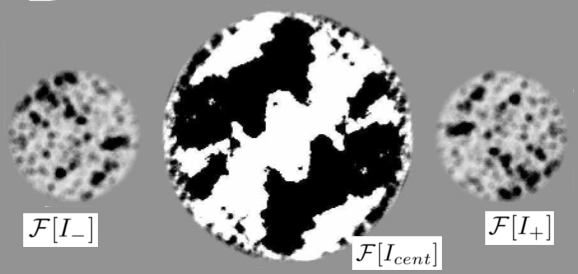}
 
   \end{tabular}
   \end{center}
   \caption[TF_of_I] 
   { \label{fig:TF_of_I} 
Fourier transform of the focal plane image with three distinct correlation peaks. We assume that these peaks are the Fourier transforms ($\mathcal{F}$ on the figure) of three images ($I_{-}$, $I_{+}$ and $I_{cent}$) summed in the focal plane}
   \end{figure}  

Under one assumption on the size and position of the reference pupil that will be described in Section \ref{II3}, we get three distinct peaks in the Fourier transform of such an image, as shown in Figure \ref{fig:TF_of_I}. We will thus express our focal plane image as the sum of three images: $I_{-}$, $I_{+}$ and $I_{cent}$, with distinct positions in the Fourier transform of the sum. The Fourier transform of one of these lateral peaks recentered can be written:

\begin{equation}
	\label{eq:I-_in_f_plan}
\ I_{-}(\alpha) = A_{S}(\alpha)A_{R}^*(\alpha).
	\end{equation}
	
Assuming a narrow spectral bandwidth, $\psi_{S}$, the complex amplitude in the pupil plane downstream of the coronagraph  can be written as: 

\begin{equation}
	\label{eq:psiS_in_p_pla}
\ \psi_{S}(u) = \mathcal{F}^{-1}[A_{S}](u) = \mathcal{F}^{-1}\left[\frac{I_{-}}{A_{R}^*}\right](u),
	\end{equation}
where $u$ is the coordinate in the pupil plane and $\mathcal{F}^{-1}$ is the inverse Fourier transform. Using $\psi_{S}$, Galicher et al. 2010\cite{Galicher10} demonstrated that we can build a simple linear estimator of the initial phase upstream of the coronagraph, assuming small aberrations. Finally, we need to estimate $A_{R}^*$, the Fourier Transform of the complex amplitude of the electric field in the reference pupil, to have a good estimation of $\psi_{S}$. We will take the assumption that $A_{R}^*$ is constant on the corrected zone. This implies that the light diffracted by the coronagraphic mask in the pupil plane downstream of the coronagraph has a uniform amplitude and phase distribution on the reference pupil. Above all, this also implies that we can consider that the reference pupil's point spread function (PSF) in the focal plane (whose size is in inverse proportion to the reference pupil size) is large enough compared to the size of the corrected zone. More accurately, we want the first zeros of the PSF to be at least outside of the corrected zone. We will call ``number of actuators", the ratio between the entrance pupil diameter and the inter-actuators distance. For a number of actuators of $N$, the corrected zone is a square area of size $N\lambda_{0}/D_{L}$x$N\lambda_{0}/D_{L}$ \cite{BordeTraub06} where $D_{L}$ is the diameter of the Lyot pupil (such a square zone is represented on Figure \ref{fig:Correctionref05}). Indeed, $N\lambda_{0}/(2D_{L})$ corresponds to the maximum frequency achievable by the $N$ actuators in each direction. Thus, the highest frequency reachable by the deformable mirror is $\sqrt{2}N\lambda_{0}/(2D_{L})$. The first dark ring of the PSF of the reference pupil is located at $1.22 \lambda_{0}/D_{R}$, with $D_{R}$, the diameter of the reference pupil. From now on, we will call $\gamma$ the ratio of the radius of the Lyot pupil versus the radius of the reference pupil ($\gamma = D_{L}/D_{R}> 1$). We will therefore consider that we are satisfying the second hypothesis if: 

\begin{equation}
	\label{eq:airy_in_dh}
\ \ 1.22\gamma > N/\sqrt{2}.
	\end{equation}

Under this assumption our estimation of $\psi_{S}$ can be written: 

\begin{equation}
	\label{eq:estim_psiS}
\ \psi_{S,estim}(u) = \mathcal{F}^{-1}\left[\frac{I_{-}(\alpha)}{A_{0}}\right](u),
	\end{equation}
	
We will now theoretically analyze the effects of the size of the reference pupil on the performances of the instrument.

\subsection{Theoretical influence of the size for small reference pupils and motivation of the study} 
\label{II3}

The size of the reference pupil will influence the quality of the correction in different ways. First of all, there is a minimum value of the distance between the centers of the two pupils, $\xi_{0} $ which depends of their size. Indeed, the three peaks in the Fourier plane (as shown in Figure \ref{fig:TF_of_I}) must be separated enough from each other, in order to select one of them properly. Using Equation \ref{eq:I_in_f_plan}, we can write this minimum separation as\cite{Galicher10}:

\begin{equation}
	\label{eq:min_sep}
\ \xi_{0} > \frac{D_{L}}{2} \left(3+\frac{1}{\gamma} \right).
	\end{equation}

\begin{figure}
   \begin{center}
   \begin{tabular}{c}
   \includegraphics[width=10cm]{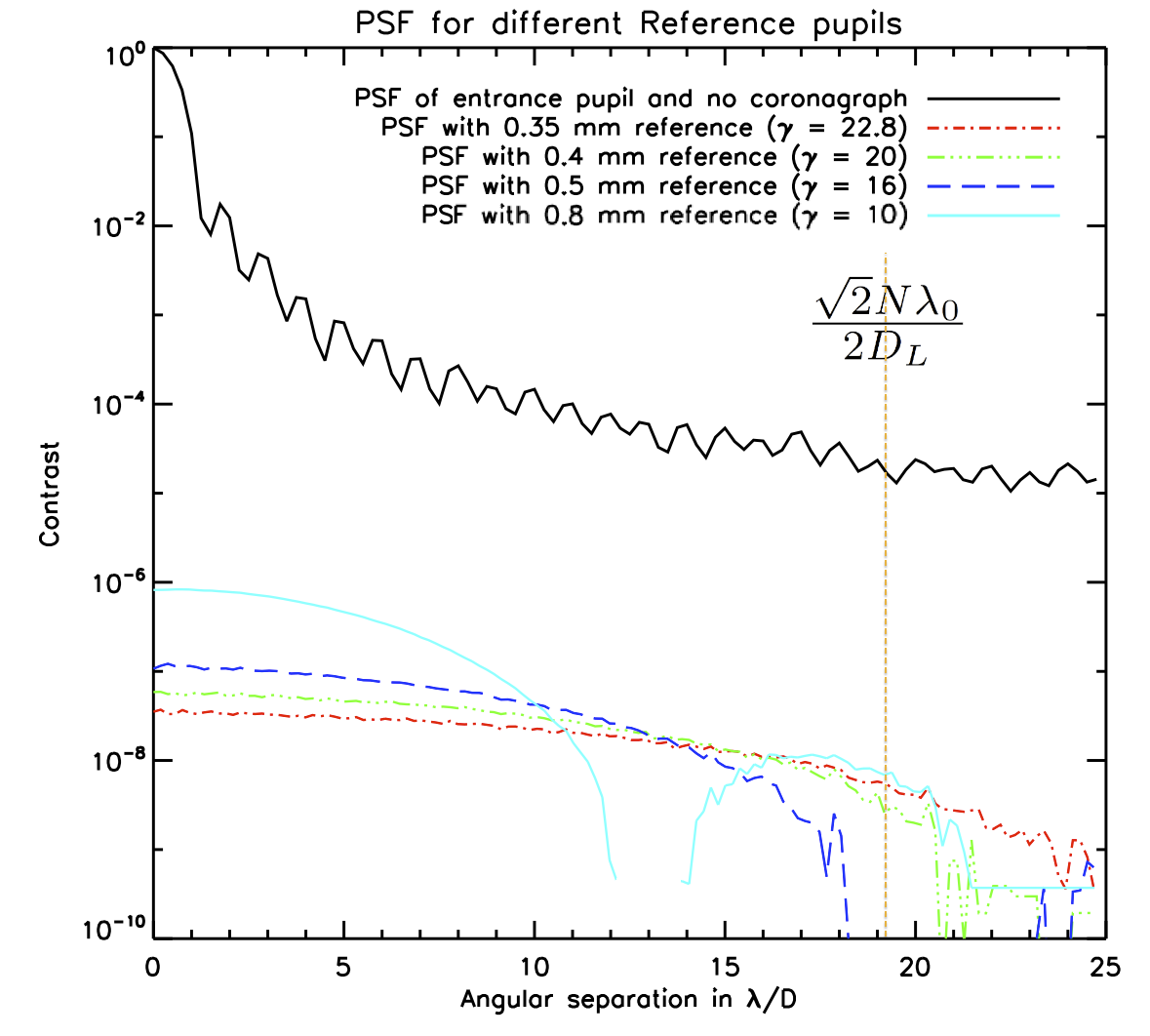}
   \end{tabular}
   \end{center}
   \caption[psf_refs] 
   { \label{fig:psf_refs} 
Radial profile of the PSF for reference pupils from $\gamma=10$ to $\gamma=22.8$ recorded on the test bench of the instrument.The distance to the center is in $\lambda_{0}/D_{L}$. These PSF are normalized by the one obtain with the instrument without coronagraph (in solid line on the figure). The vertical line correspond to the limit of the correction zone for a $N = 27$ actuators in the entrance pupil ($\sqrt{2}N\lambda_{0}/(2D_{L})$)}
   \end{figure}  

A second effect is the additional luminosity which is added to the focal plane by the reference pupil. Figure \ref{fig:psf_refs} shows the recorded radial profiles of the focal plane images with different reference pupil's diameters, normalized to the maximum intensity of the PSF through Lyot pupil without coronagraph (black, solid in Figure \ref{fig:psf_refs}). The intensity level varies from $10^{-6}$ to $10^{-8}$ for decreasing reference pupil's diameters (from $\gamma=10$ to $\gamma=22.8$). If we want to use the SCC down to contrasts of $10^{-8}$ or better, as in Baudoz et al. 2012\cite{BaudozSPIE12}, the light issue of the reference pupil may limit the performances of the SCC. One solution is to correct the aberrations using the SCC then hide the reference pupil to record the image in coronagraphic mode. However, this solution does not allow the correction in real time. In all cases, we understand that a small reference pupil is preferable. 

We are now going to analyze the two main effects of the size of the reference pupil on the SCC performances :

\begin{enumerate}
\item The size of the reference pupil has an influence on the luminosity of the fringes. The larger the reference pupil is the brighter the fringes will be. If the contrast of the fringes (the difference between dark and bright fringes) is smaller than the noise (camera and photon noise), $I_{-}$ will not be accessible and therefore the correction will be impossible. In our case, the correction is limited by the camera noise. This effect advocates for larger reference pupils

\item In Section \ref{II2}, we made the assumption that $A_{R}^*$ is constant over the all the correction zone. This is equivalent to saying that variations in the luminosity of the fringes are exclusively due to aberrations. But this assumptions is better satisfied with very small reference pupils because their PSF in the focal plane are significantly larger than the corrected zone. Our estimator will thus be more accurate in the case of small reference pupil, which will lead to a better correction

\end{enumerate}

We can predict that the first effect will be more critical because it varies with the luminosity of the reference pupil and thus the square of its radius, while the second one varies with the size of the PSF, which is linearly proportional to the radius of the reference pupil.

A theoretical analysis of the influence of the size of the reference pupil on the performances of the instrument reveals two main and opposing effects. It also indicates why a good knowledge of these effects is required to use the instrument optimally. Even if the tests show that it is preferable to use very small reference pupils, very large ones can also be very useful. In fact, some cases (many aberrations due to an unknown initial position of the deformable mirror, for example) require the use of very large reference pupils which will produce fringes even with very aberrated wave front and will help us to initiate the correction, to maybe use a smaller reference pupil afterward.

\subsection{Bench description} 
\label{II4}
The test bench is described in details in Mas et al. 2010\cite{mmas_spie_2010}. We recall the main components below.

\begin{enumerate}
\item A laser diode emitting at 635nm with a quasi-monochromatic bandwidth
\item A tip-tilt mirror and a Boston Micromachines deformable mirror of 32x32 actuators (only 27x27 used)
\item A coronagraph including:
\begin{itemize}
\item[(a)] A Four quadrant phase mask\cite{Rouan00} (FQPM) optimized for a wavelength of 635nm
\item[(b)] A Lyot stop with a size of 8mm for an entrance pupil of 8.1mm
\item[(c)] In the same plan, a reference pupil of variable size: 0.3mm ($\gamma = 26.6$), 0.35mm ($\gamma = 22.8$), 0.4mm ($\gamma = 20$), 0.5mm ($\gamma = 16$), 0.8mm ($\gamma = 10$) and 1.5mm ($\gamma = 5.3$)
\end{itemize}
\item A CCD camera of 640 by 480 pixels of which we use 400 by 400 with a readout noise of 18 electrons and a full well capacity of 13,000 electrons
\item A set of neutral density filters used to record unsaturated non-coronagraphic images which help us to normalize image intensities
\item Software (Labview) to control the deformable mirror, build the interaction matrix and close the control loop that will allow real time correction. For more information about the construction of the synthesis matrix, see Mas et al.2012\cite{MasSPIE12}
\end{enumerate}

\section{Simulation and experimental study of the influence of the reference pupil}
\label{III}
For our correction, we are not using all of the 32x32 actuators of our deformable mirror. The number of actuators N in the entrance pupil (that we defined as the ratio between the entrance pupil diameter and the inter-actuators distance) is 27. Thus, our largest corrected zone is $27\lambda_{0}/D_{L}$x$27\lambda_{0}/D_{L}$. Using Equation \ref{eq:airy_in_dh} with $N = 27$, we can separate the study of our reference pupils into two sets of cases. The ones which we will call ``small reference pupils", with $\gamma > 15.5$, and for which the first dark ring of the PSF of the reference pupil is larger than the corrected zone (the one with $\gamma > 16$ thus barely satisfies this condition), and the ones which we will call ``large reference pupils", where this ring is inside the corrected zone. We proceeded to series of tests using the different reference pupils.

It is possible to reduce the corrected zone by choosing to multiply $I_{-}$ by a smaller square mask. If we do so, Equation \ref{eq:estim_psiS}, which gives an estimator of the complex amplitude in the pupil plane downstream of the coronagraph $\psi_{S}$, can now be written:

\begin{figure}
   \begin{center}
   \begin{tabular}{c}
   \includegraphics[width=5.5cm]{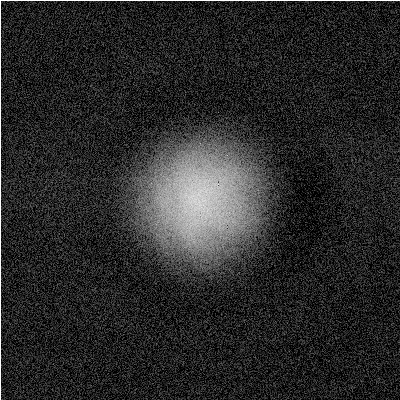}
   \includegraphics[width=5.5cm]{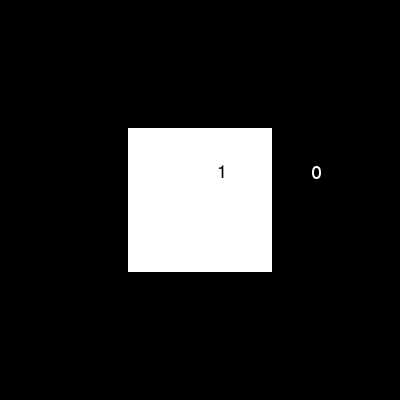}
   \includegraphics[width=5.5cm]{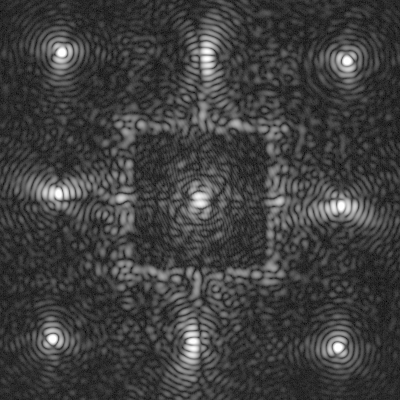}
   
   \end{tabular}
   \end{center}
   \caption[Correctionref05] 
   { \label{fig:Correctionref05} 
PSF of the 0.5mm reference pupil ($\gamma = 16$) (right). As this PSF is just barely larger than the correction zone, we construct a square zone of $25\lambda_{0}/D_{L}$ (center), which we call $Sq(\alpha)$ in Equation \ref{eq:psiS_in_p_plan_modif}. We multiply $I_{-}$ by this mask to achieved the correction (right).}
   \end{figure}  

\begin{equation}
	\label{eq:psiS_in_p_plan_modif}
\ \psi_{S,estim}(u) = \mathcal{F}^{-1}\left[\frac{I_{-}(\alpha)Sq(\alpha)}{A_{0}}\right](u).
	\end{equation}

Where $Sq(\alpha)$ equals 1 on a square area of $N'\lambda_{0}/D_{L}$x$N'\lambda_{0}/D_{L}$ in the center of the image and 0 everywhere else. Such a mask is represented on Figure \ref{fig:Correctionref05} (center). If we take $N'\leq27$, the corrected zone is smaller than the maximum one we can achieve, but there are multiple advantages. Firstly, as explained in Poyneer et al. 2004\cite{Poyneer04} the spatial frequencies that the mirror cannot correct (those which are higher than $27/2\lambda_{0}/D_{L}$) introduce aliasing in the estimation of the phase defects. This method thus allows a better correction where $Sq(\alpha)$ equals 1. Secondly, by reducing the $Sq(\alpha)$, the second effect described in Section \ref{II3} will be less important: for the same reference pupil, the assumptions of a constant PSF over the correction zone will be more satisfied.

\subsection{Illustration of the effects of the size of reference pupil by simulation} 
\label{III1}

\begin{figure}
   \begin{center}
    \begin{tabular}{c}
   \includegraphics[width=8.2cm]{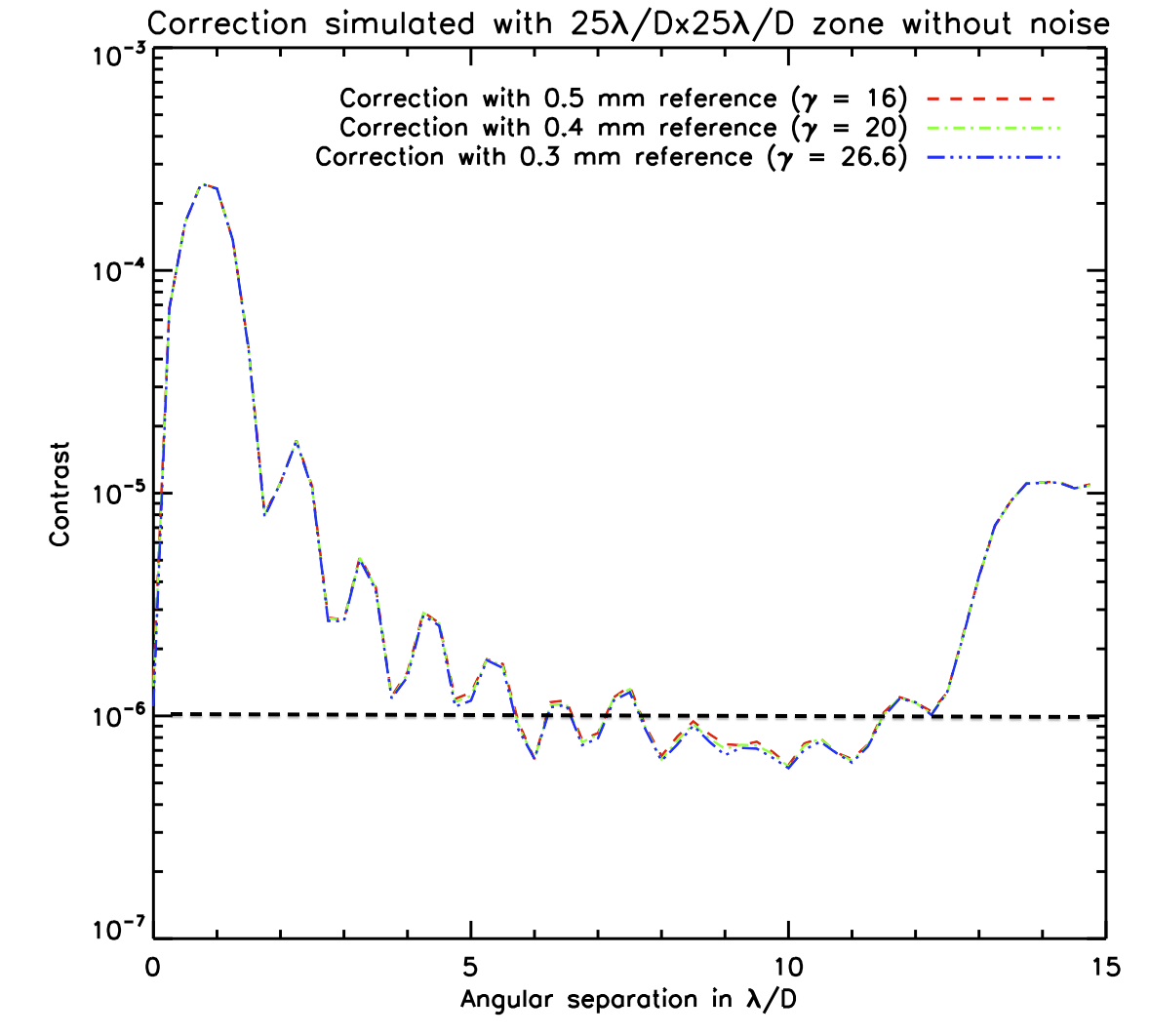}
   \includegraphics[width=8.2cm]{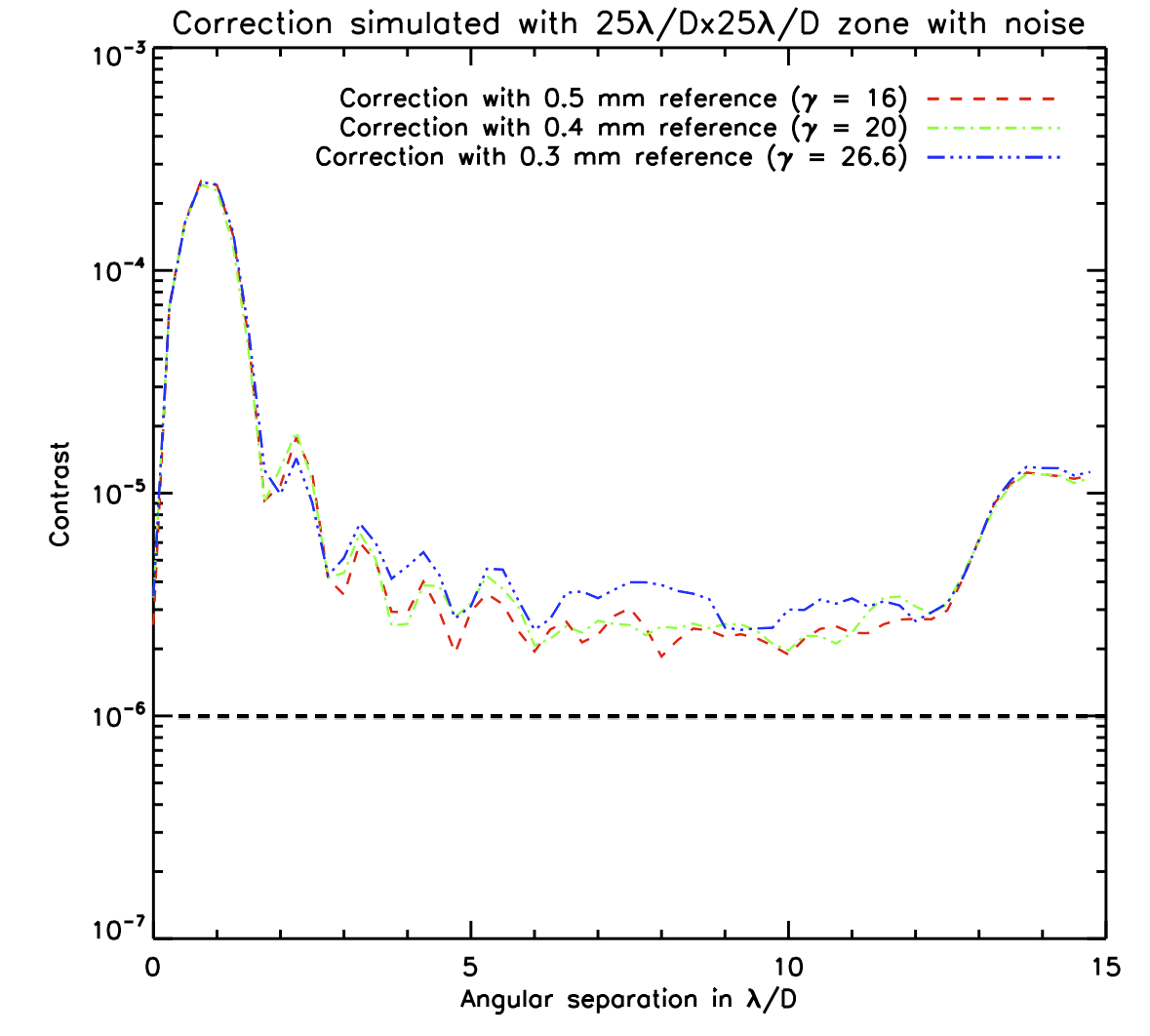}
   \end{tabular}
   \end{center}
   \caption[simu_effectsbruit] 
   { \label{fig:simu_effectsbruit} 
Radial profiles of the simulation results for two different reference pupils in a small dark zone ($25\lambda_{0}/D_{L}$x$25\lambda_{0}/D_{L}$) without (left) of with (right) noise simulated. These mean intensities are normalized by the maximum of the PSF obtained without coronagraph}
   \end{figure}  

\begin{figure}
   \begin{center}
   \begin{tabular}{c}
   \includegraphics[width=8.2cm]{testa3ptitezone_ssbruit.png}
   \includegraphics[width=8.2cm]{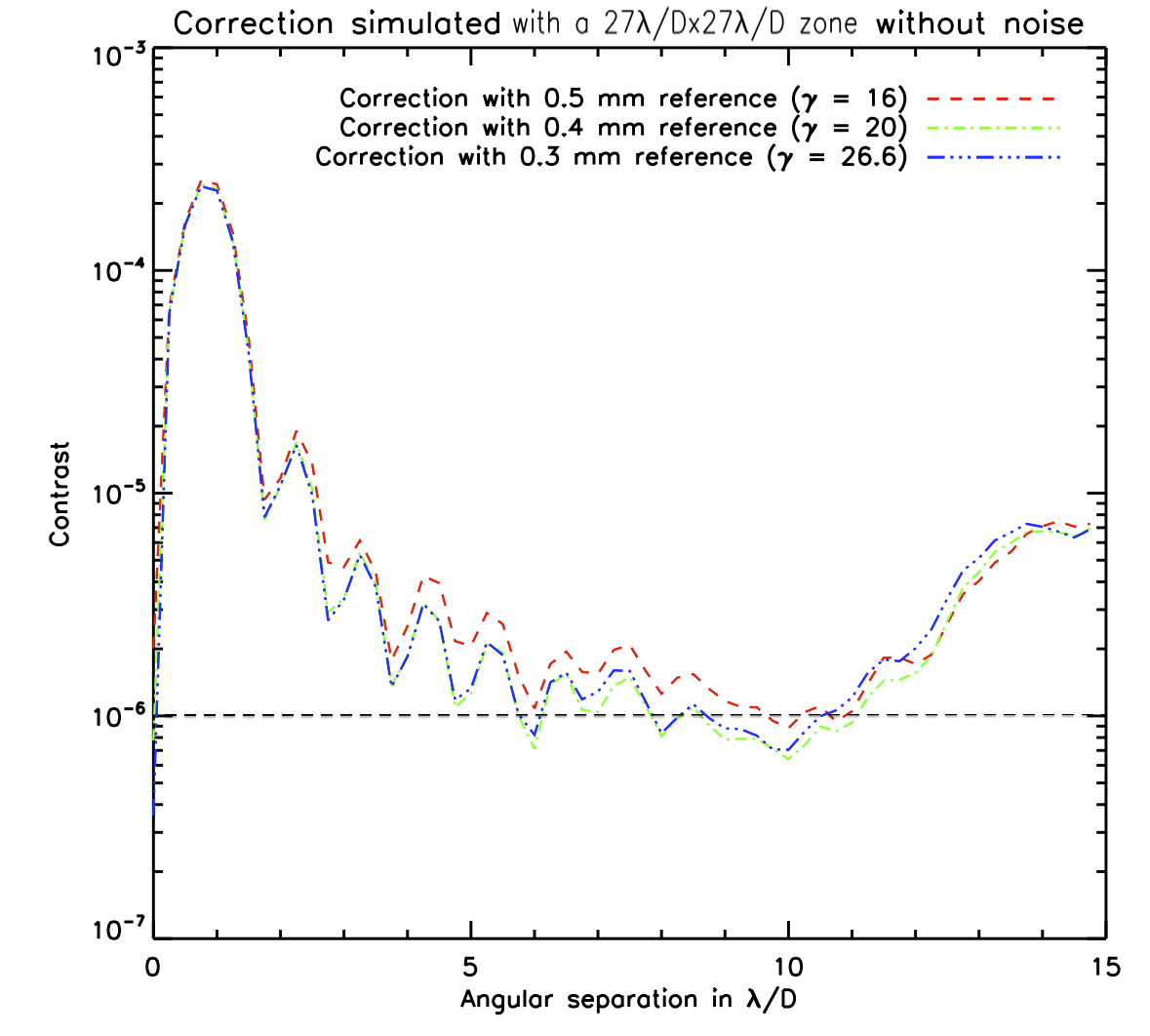}
   \end{tabular}
   \end{center}
   \caption[simu_effectsPSF] 
   { \label{fig:simu_effectsPSF} 
Radial profiles of the simulation results for two different reference pupils in a smaller dark zone ($25\lambda_{0}/D_{L}$x$25\lambda_{0}/D_{L}$) (left),or with all the zone simulated. We did not simulate any noise. All the mean intensities are normalized by the maximum of the PSF obtained without coronagraph}
   \end{figure}

Using simulation tools, we can separate the two different effects formulated in Section \ref{II3} (noise and size of the reference pupil's PSF) and analyze their influence on the performances of the instrument. The simulation results presented in this paper have been achieved with an upgrade of the software of R. Galicher called MEHICI (for Meudon High Contrast Imaging code). In these simulations, we assumed phase aberrations of 30nm root mean square (RMS), with a power spectral density (PSD) in $f^{-2}$. To match simulation results with recorded data, we must introduce realistic amplitude defects. Therefore, we use an image of the pupil illumination recorded on the bench to estimate these amplitude defects. Since they are rather large (about
 $9 \% $ RMS in intensity), the results of the simulation (Figure \ref{fig:simu_effectsPSF} and Figure \ref{fig:simu_effectsbruit}) as well as the bench tests (Figure \ref{fig:test_bench}) show a very poor correction (very low contrasts) near the center (between 0 $\lambda_{0}/D_{L}$ and 4 $\lambda_{0}/D_{L}$). Although, the SCC can estimate amplitude defects and allow their correction (on half of the correction zone, as we are only using one deformable mirror in our test bench\cite{BordeTraub06}), we do not study this capability in this paper. More information on amplitude correction with SCC can be found in Mas et al.2012\cite{MasSPIE12} and Baudoz et al. 2012 \cite{BaudozSPIE12}.

First, we study the effect of camera and photon noise on the SCC performances. The curves on Figure \ref{fig:simu_effectsbruit} are the mean contrast of the pixels in the image at a constant angular distant of the center given on the x-axis in $\lambda_{0}/D_{L}$. We simulated the SCC correction for three reference pupils ($\gamma = 16$, $\gamma = 20$ and $\gamma = 26.6$) and use a a correction zone $Sq(\alpha)$ of $25\lambda_{0}/D_{L}$x$25\lambda_{0}/D_{L}$. A camera noise can be added to the simulated image to study its effect on the SCC control loop: we compare these mean radial contrast in the image without noise (left) and with noise (right). The camera noise normalized to the maximum intensity of the image without coronagraph is $3.10^{-7}$ in the test bench described in Section \ref{III2}. However, to clearly show the effect of the noise, we deliberately simulated a stronger noise: $9.10^{-7}$. Because we are using a correction zone smaller than any of the PSF, the effect of the size of the PSF is minimized and as we are not simulating any noise on the left figure the radial profile shows no difference between the three reference pupils. We see clearly that when we add simulated noise, the larger the reference pupil is, the more efficiently the SCC corrects. This was expected from the first effect described in Section \ref{II3}: with a larger reference pupil, the reference pupil's flux is more important, and the fringes will be more contrasted. 

Simulation also allows to study the second effect described in Section \ref{II3} (approximation of the PSF to a constant), as shown on Figure \ref{fig:simu_effectsPSF}. We are simulating noise-free simulations with the same reference pupils ($\gamma = 16$, $\gamma = 20$ and $\gamma = 26.6$) than in Figure \ref{fig:simu_effectsbruit}. Figure \ref{fig:simu_effectsPSF} (left) is the radial mean contrast of the results obtain with a simulation realized with a correction zone $Sq(\alpha)$ of $25\lambda_{0}/D_{L}$x$25\lambda_{0}/D_{L}$, while Figure \ref{fig:simu_effectsPSF} (right) is realized with an $Sq(\alpha)$ of $27\lambda_{0}/D_{L}$x$27\lambda_{0}/D_{L}$. As the $25\lambda_{0}/D_{L}$x$25\lambda_{0}/D_{L}$ correction zone is much smaller than any of the reference pupil's PSF, the effect of the size of the PSF is minimized and as we are not simulating any noise on the left figure the radial profile shows no difference between the three reference pupils. But the largest reference pupil ($\gamma = 16$) has a PSF barely larger than the $27\lambda_{0}/D_{L}$x$27\lambda_{0}/D_{L}$ corrected zone: for this $\gamma$ and this correction zone, the contrast achieved is poorer than for the two smaller sizes, as shown on the right figure. This is due to the fact that  our phase estimation for this reference pupil's size is inaccurate because the assumption of a constant reference PSF over the correction zone is not satisfied. 

This simulation study allowed us to isolate the two effects described in Section \ref{II3}. We are now going to present the results obtained on the test bench for different reference pupil's sizes.

\subsection{Bench result for small reference pupils (high gammas)} 
\label{III2}

\begin{figure}
   \begin{center}
    \begin{tabular}{c}
   \includegraphics[width=8.2cm]{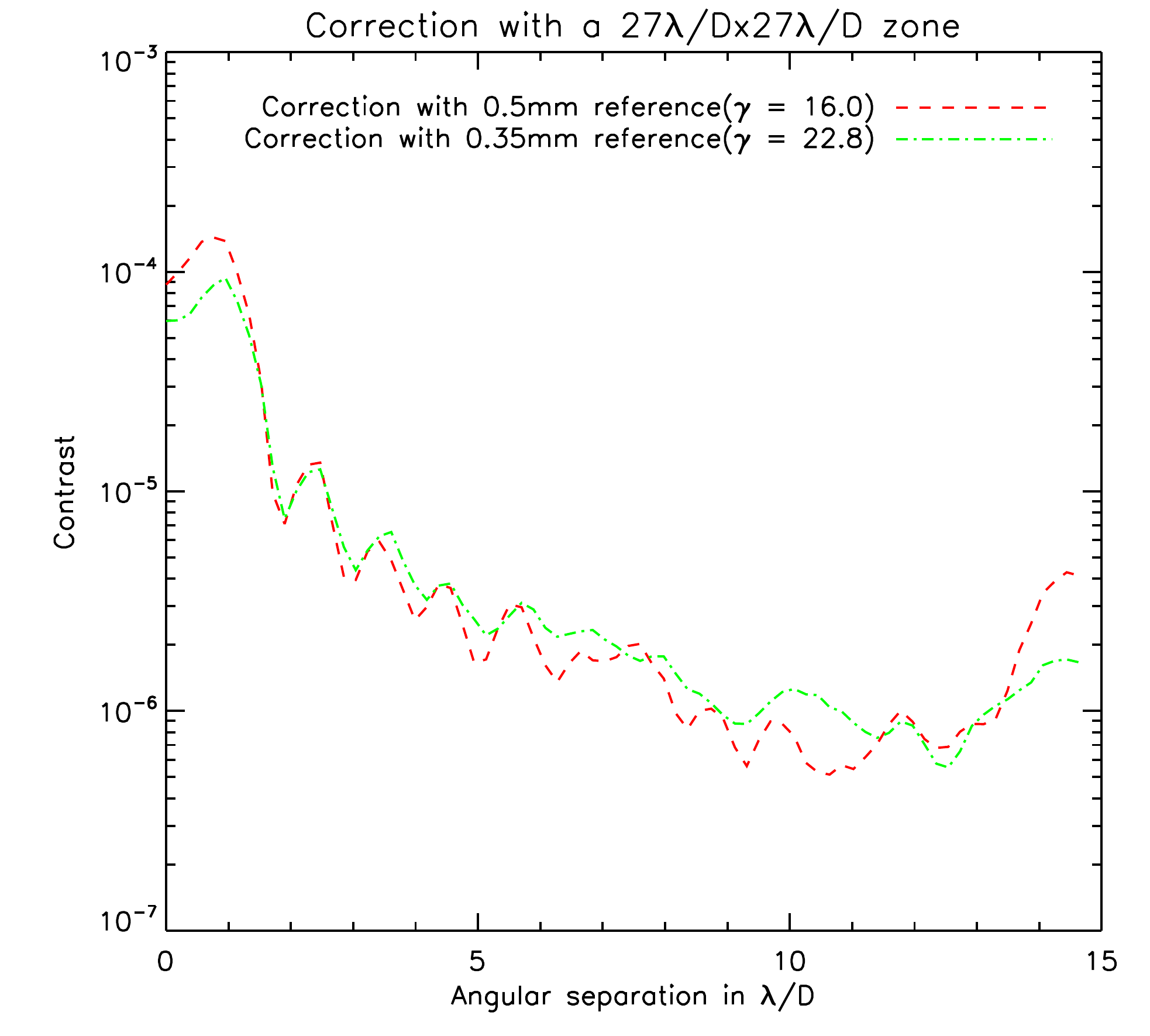}
   \includegraphics[width=8.2cm]{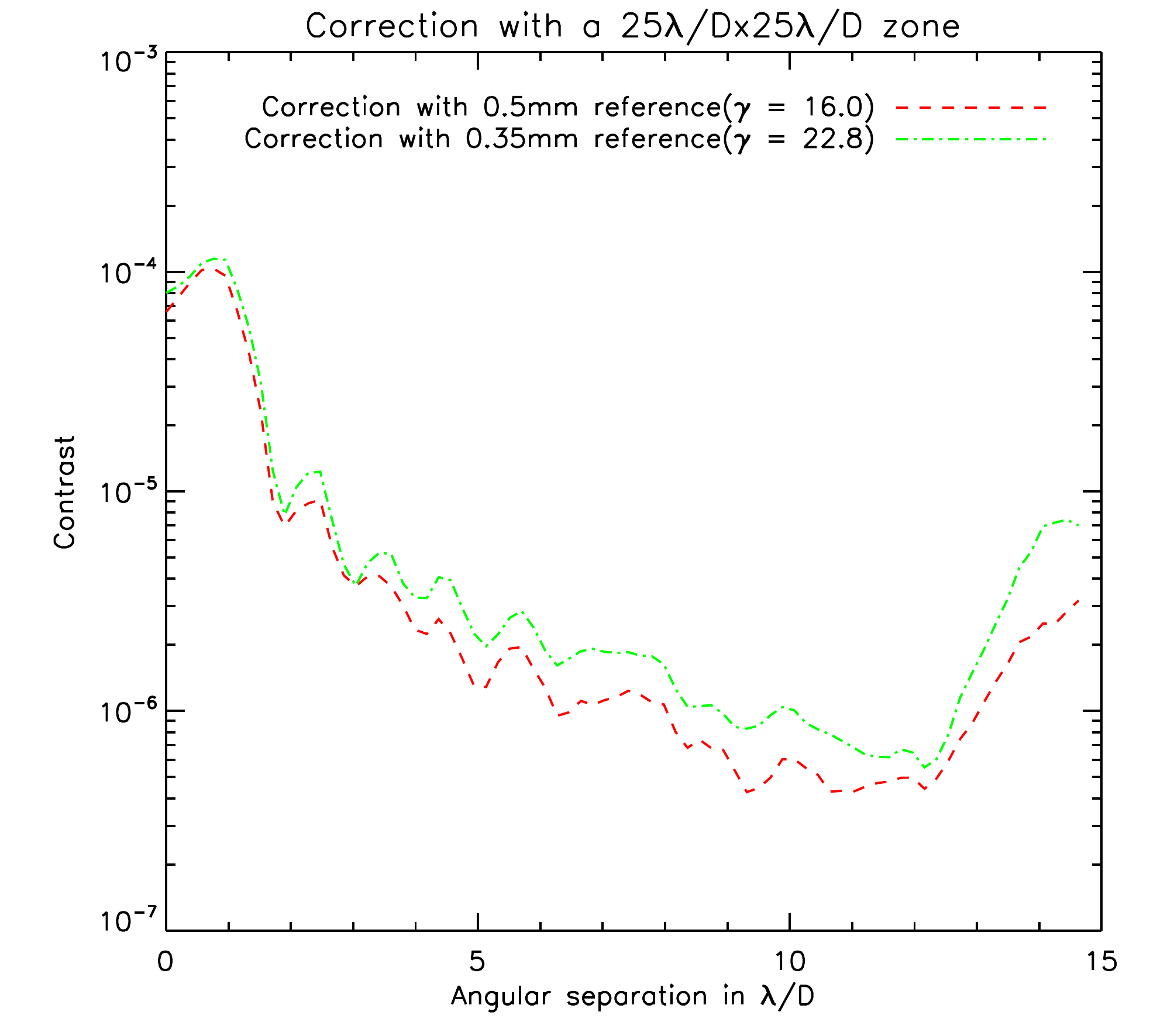}
   \end{tabular}
   \end{center}
   \caption[test_bench] 
   { \label{fig:test_bench} 
Radial profiles obtained on the test bench for two different reference pupils, with full size correction zone (left) or with a slightly smaller zone (right). These mean intensities are normalized by the maximum of the PSF obtained without coronagraph nor SCC}
      \end{figure}

We tested both corrected zone $Sq(\alpha)$ of $27\lambda_{0}/D_{L}$x$27\lambda_{0}/D_{L}$ (radial profile of the results presented on Figure \ref{fig:test_bench}, left) and $25\lambda_{0}/D_{L}$x$25\lambda_{0}/D_{L}$ (Figure \ref{fig:test_bench}, right) with $\gamma = 16$ and $\gamma = 22.8$. If we decided to use the smaller corrected zone $Sq(\alpha)$ of $25\lambda_{0}/D_{L}$x$25\lambda_{0}/D_{L}$ (Figure \ref{fig:test_bench}, right), the contrast obtained with the largest reference ($\gamma = 16$) pupil is twice better than the one obtained with the smallest one ($\gamma = 22.8$). If we use the largest $Sq(\alpha)$ ($27\lambda_{0}/D_{L}$x$27\lambda_{0}/D_{L}$ zone, left on Figure \ref{fig:test_bench}), the largest reference pupil ($\gamma = 16$) is still slightly better (except maybe for the speckles close to the edges of the dark hole). This tends to show that the effect of the size of the PSF is not that critical compare to the effect of the noise (which we estimated at $3.10^{-7}$ for these tests) because even with a reference pupil that barely satisfying Equation \ref{eq:airy_in_dh}, we reach a better correction. 

\subsection{Bench result for large reference pupils (small gammas)} 
\label{III3}

\begin{figure}
   \begin{center}
   \begin{tabular}{c}
   \includegraphics[width=5.5cm]{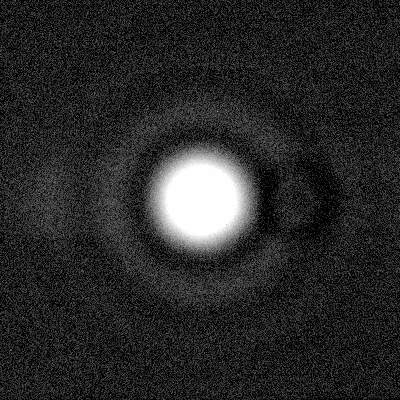}
   \includegraphics[width=5.5cm]{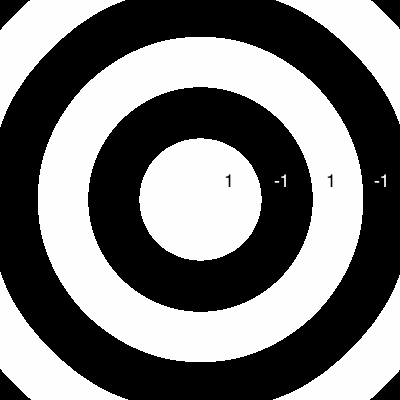}
   \includegraphics[width=5.5cm]{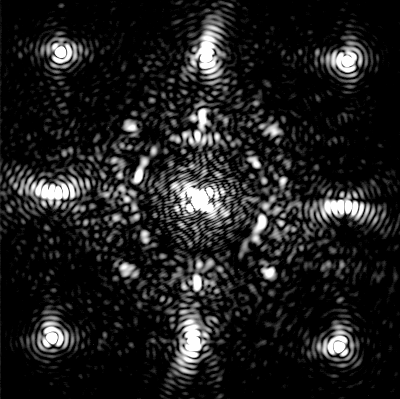}
   
   \end{tabular}
   \end{center}
   \caption[Correctionref08] 
   { \label{fig:Correctionref08} 
PSF of the 0.8mm reference pupil ($\gamma = 10$) (right). From this PSF we construct the sign mask (center). Multiplying $I_{-}$ by this mask and the correction can be achieved (right) for this reference pupil.}
   \end{figure}

With very large reference pupils (when Equation \ref{eq:airy_in_dh} is not satisfied, and the PSF is actually smaller than the corrected zone covered by the mirror), we must modify the phase estimation, because the assumptions of a constant reference pupil in the corrected zone is not at all satisfied: the first dark ring of the PSF (shown in Figure \ref{fig:Correctionref08}, left) is in the correction zone. This dark rings correspond to a zone where $A_{R}$ is null and thus, the speckles in this region are not fringed, which make the estimation impossible. Therefore, we cannot correct the speckles in this region. Beyond this ring and before the next one (at $2.23 \lambda_{0}/D_{L}$), the sign of $A_{R}$ changes. We are going to use another assumption: instead of $A_{R}$, $|A_{R}|$ is now constant on the corrected zone. Our estimator of $\psi_{S}(u)$ now becomes: 

\begin{equation}
	\label{eq:psiS_in_p_plan_modif_sgn}
\ \psi_{S,estim}(u) = \mathcal{F}^{-1}\left[\frac{I_{-}(\alpha).Sign[\Re[A_{R}^*]](\alpha)}{A_{0}}\right](u),
	\end{equation}
	
where $Sign[\Re[A_{R}^*]]$, is the sign of the real part of $A_{R}^*$.
To achieve the correction with the 0.8mm ($\gamma = 10$) reference pupil we multiply $I_{-}$ by the mask showed in Figure \ref{fig:Correctionref08} (center), where the white zone values are equal to $1$ and the black zone one equals to $-1$. Figure \ref{fig:Correctionref08} (right) shows the results obtained on the bench. We can distinctly see the first ring at $1.22 \lambda_{0}/D_{R}$. As expected, this ring is not corrected, because the speckles are not fringed where the reference pupil's PSF intensity is null. Although we know that correction with a very large reference pupil is possible, the level of speckles suppression is much lower (lower contrasts) than with the other reference pupils, presumably because the uncorrected frequencies of the ring have an influence on the corrected zone. Incidentally, our attempts to achieve a correction with the largest reference pupil of 1.5mm ($\gamma = 5.3$) only produced a very weak correction inside the first dark ring and no correction at larger angular separation.

\section{Conclusion}

Using simulation tools, we identified two main effects of the size of the reference pupil on the performances of the SCC: 
\begin{itemize}
\item The noise (camera and photo noise) advantages large reference pupils because they produce brighter fringes that have a higher signal to noise ratio
\item A small PSF (i.e. a large reference pupil) corrects poorly the speckles on the edges of the corrected zone
\end{itemize} 
The results on the bench show that for the moment, the noise effect is more critical. But the camera on the test bench is about to be replaced by a sCMOS camera with much less noise (1.5e- noise instead of 18e-) which will lead to better performances of the smallest reference pupils. Our best results in contrast, as shown in Figure \ref{fig:test_bench}, are better than $10^{-6}$ between 5 and 12.5 $\lambda_{0}/D_{R}$ for the moment. We recall that we did not use amplitude correction in this study and that the correction under 5$\lambda_{0}/D_{R}$ is limited by amplitude defects. A new correction method using the same instrumental design of the SCC is showing much better results: a few $10^{-8}$ (see Baudoz et al. 2012\cite{BaudozSPIE12} for more information about this technique). In this case, the fluxes of the reference pupil (see Figure \ref{fig:psf_refs}) will become a problem and smaller reference pupils will be advantaged.

\bibliography{report}   
\bibliographystyle{spiebib}   

\end{document}